\documentclass[twocolumn]{aastex631}

\usepackage{graphicx}
\usepackage{dcolumn}
\usepackage{bm}
\usepackage{xcolor}
\usepackage{hyperref}
\usepackage{amsmath}
\usepackage[mathlines]{lineno}
\newcommand{\mone}{m_1}
\newcommand{\mtwo}{m_2}
\newcommand{\chieff}{\chi_{\rm eff}}
\newcommand{\chip}{\chi_p}
\newcommand{\Msun}{M_\odot}
\newcommand{\GWTCfive}{GWTC–5.0}
\newcommand{\bbh}{BBH}

\newcommand{\dd}{\mathrm{d}}

\newcommand{\pistroke}{
\text{\protect\ooalign{\hidewidth\raisebox{-0.2ex}{--}\hidewidth\cr$\pi$\cr}}}
\newcommand{\tn}{\ensuremath{\mathcal{N}_t}}

\newcommand{\popL}{\texttt{low-mass}}
\newcommand{\popMH}{\texttt{horizontal}}
\newcommand{\popMD}{\texttt{diagonal}}
\newcommand{\popU}{\texttt{high-mass}}

\newcommand{\pL}{p_{\popL}}

\newcommand{\pMD}{p_{\popMD}}
\newcommand{\pU}{p_{\popU}}

\newcommand{\paperrepo}{\url{https://github.com/NirGutt/Four-Subpopulations-of-Binary-Black-Hole-Mergers}}

\newcommand{\SPA}{School of Physics and Astronomy, Monash University, Vic 3800, Australia}
\newcommand{\OzGravMonash}{OzGrav: The ARC Centre of Excellence for Gravitational Wave Discovery, Clayton VIC 3800, Australia}

\begin{document}
%\linenumbers

\title{Revealing Four Subpopulations of Binary Black-Hole Mergers \\
with the Fifth Gravitational-Wave Transient Catalog}

\author{Nir Guttman}
\email{nir.guttman@monash.edu}
\affiliation{\SPA}
\affiliation{\OzGravMonash}

\author{Paul D. Lasky}
\affiliation{\SPA}
\affiliation{\OzGravMonash}

\author{Eric Thrane}
\affiliation{\SPA}
\affiliation{\OzGravMonash}

\begin{abstract}
Gravitational-wave data are beginning to reveal a structured landscape of black-hole masses and spins, suggesting multiple formation processes are now being resolved observationally. We analyze data from LIGO--Virgo--KAGRA's (LVK's) fifth Gravitational-Wave Transient Catalog and find that the population is naturally described by four distinct subpopulations. The dominant component, contributing $\simeq70\%$ of the astrophysical merger rate, is characterised by a low-mass population centred near $10\,\Msun$ and is separated from heavier systems by a depletion near $14\,\Msun$. 
This component may be associated with black holes formed from failed supernovae.
Above this depletion, we find two intermediate-mass components: an unequal-mass branch pairing the lower-mass, $\simeq10\,\Msun$ black hole with a heavier black hole, perhaps associated with isolated-binary/stable-mass-transfer formation, and a nearly equal-mass branch peaking near $30$--$35\,\Msun$ whose low spins and mass distribution favour first-generation systems possibly born in dense stellar environments. A fourth, percent-level component extends to higher masses and is characterized by a broad mass-ratio distribution and large spin magnitudes, consistent with a hierarchical-merger population. Our four-component model is overwhelmingly preferred over a standard LVK population model by a natural-log Bayes factor of $\ln\mathcal{B}=19.2$. 
Our work observationally unveils a new subpopulation of black-hole mergers utilising a new hybrid data-driven and parametric-model discovery method, bringing us one step closer to understanding stellar-mass black-hole archaeology.

\end{abstract}

\section{Introduction}
The growing catalog of gravitational-wave observations is revealing increasingly detailed patterns in the binary-black-hole (\bbh) population. Structure continues to emerge across  the mass, mass-ratio, redshift, and spin distributions.
The fifth Gravitational-Wave Transient Catalog~\cite[\GWTCfive;][]{gwtc-5} population analysis reports evidence for a peak in the merger rate near $10\,\Msun$, a change in slope around $35\,\Msun$, preferential equal-mass pairing near this feature, unequal-mass systems at high primary masses, and spin distributions that vary with mass and mass ratio~\citep{gwtc-5_population}. 
These results extend earlier indications~\citep{gwtc3_pop,gwtc4_pop} of substructure in the \bbh{} mass spectrum and strengthen the case for a population with multiple astrophysical components.

Recent work has sharpened this picture. Analyses of \GWTCfive{} and earlier catalogs find that the distributions of both the mass ratio $q$ and the effective inspiral spin parameter $\chi_\text{eff}$ undergo abrupt changes at certain values of primary mass $m_1$, suggesting that different regions of the mass spectrum are associated with distinct pairing and spin properties~\citep{Talbot_2018,Mould_2022,hui_chi_eff_gp,binned_gp_2025,pix_pop,Sridhar:2025kvi,Ray_2026a,pix_pop_gwtc5,Farah_subpopulation,Sylvia_subpopulation,Banagiri_GWTC4}. 
Complementary studies of the component-spin distributions have also found evidence that spin magnitudes and orientations vary across the black-hole mass spectrum~\citep{Li_2024,Galaudage_2025,galaudage_2026,berti_2026,wolfe2026}.
Independent studies of the high-mass population have argued that the pair-instability mass gap may be most clearly visible in the secondary-mass distribution, with primary black holes inside the gap plausibly produced by previous mergers~\citep{hui_gap}. Similarly, \citet{antonini_U_chi_eff} showed that hierarchical mergers in dense stellar environments predict broad and nearly symmetric $\chieff$ distributions above the pair-instability supernovae gap, providing a characteristic spin signature for a high-mass dynamical population. At lower masses, \citet{Failedsn_Katerina} found evidence for a gap immediately above the $10\,\Msun$ feature, which may be associated with compactness-driven variations between successful and failed supernovae~\citep{Schneider_2021,Schneider_2023,Failedsn1,Failedsn_Katerina}. Together, these studies suggest that the features seen in the mass spectrum are connected to changes in the pairing and spin distributions, pointing to multiple physical processes shaping the observed \bbh{} population.

A central challenge is that the relevant structure may not be obvious a priori. Parametric population models are powerful when the appropriate morphology is somewhat known, but they can miss unexpected features or yield misleading fits when the model is misspecified. Conversely, highly flexible reconstructions can reveal unexpected structure, but the robustness and astrophysical interpretation of that structure may not be immediately clear. We therefore adopt a two-stage strategy. First, we use $\pistroke$ (\emph{pi-stroke}), a non-parametric maximum-likelihood population-inference method~\citep{pi-stroke1,pi-stroke2}, as an exploratory tool to identify localized structure in the \bbh{} mass distribution with minimal assumptions about its morphology. The method represents the population as a weighted collection of delta functions, whose locations and weights are optimized to maximize the population likelihood. We use this reconstruction as a discovery step: it identifies where the data prefer probability density in the two-dimensional component-mass space.

Guided by the $\pistroke$ reconstruction, we then construct a parametric model for the \GWTCfive{} \bbh{} population. The model decomposes the population into a small number of subpopulations, first defined by their morphology in the $(m_1,m_2)$ plane, where $m_1$ and $m_2$ denote the primary and secondary black-hole masses, respectively. Each mass-defined subpopulation is then assigned its own spin distribution. 

This procedure reverses the usual order of inference: rather than beginning with a theoretically motivated model and searching for its imprint in the data, we first identify and characterize the dominant structures preferred by the observations and only then interpret them astrophysically. This is particularly useful given the large number of possible \bbh{} formation scenarios and the difficulty of mapping any single theoretical channel uniquely onto the observed population.

We find evidence for four \bbh{} components, or subpopulations. These include nearly equal-mass branches associated with different mass scales, an asymmetric component whose secondary mass is concentrated near the low-mass $\sim10\,\Msun$ scale, and a high-mass component extending toward the pair-instability regime. These components exhibit distinct spin behaviour, reinforcing the interpretation that the observed mass features correspond to different subpopulations rather than to a single one.

\section{Data and population model}
\label{sec:data_model}

We analyze the \GWTCfive{} \bbh{} events using the same event selection and population-inference settings as the LVK population analysis~\citep{gwtc-5_population}. We include events with false-alarm rate ${\rm FAR}\leq1\,{\rm yr}^{-1}$ and component masses satisfying $m_1,m_2\geq3\,\Msun$ at 99\% credibility, yielding 259 events. Hierarchical inference is performed with \texttt{gwpopulation}~\citep{gwpopulation}. Selection effects are included using the standard injection-based estimate of the detection efficiency, following the LVK population-analysis prescription~\citep{gwtc-5_population}.

For each event, we use posterior samples transformed to the source-frame variables $(\mone,\mtwo,z,\chi_1,\chi_2,\cos\theta_1,\cos\theta_2)$,
where $z$ is the redshift, $\chi_i$ are the dimensionless spin magnitudes, and $\theta_i$ are the spin--orbit tilt angles.

The first step in our analysis is to identify the main structures preferred by the data without imposing a specific parametric form. To do this, we use $\pistroke$, a maximum-population-likelihood reconstruction method. The $\pistroke$ distribution is defined as the population distribution that maximizes the hierarchical population likelihood. As shown by \citet{pi-stroke1}, this maximum-likelihood solution always takes the form of a weighted sum of delta functions. It should not be interpreted as the true astrophysical population distribution, nor as a smooth density estimate. Instead, it indicates where the data prefer the population model to place support.

\begin{figure*}[t]
\centering
\includegraphics[width=0.95\textwidth]{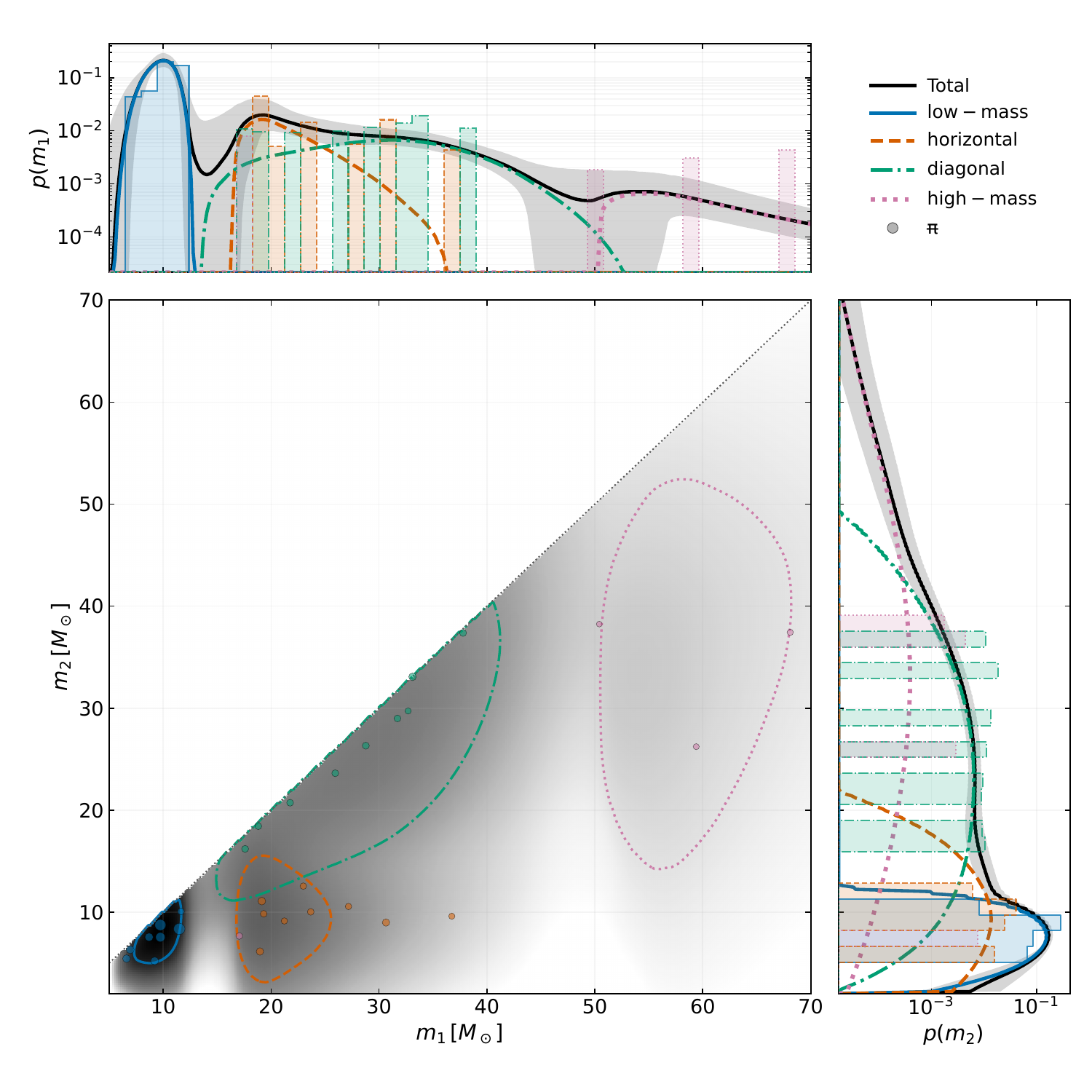}
\caption{
The distribution of black hole component masses.
The central panel shows the reconstructed population density in the $(m_1,m_2)$ plane, with grey shading denoting the total distribution.
Darker shading indicates higher probability density. 
The coloured contours show 25\% peak-density levels of different sub-populations identified in this study. 
The top and right panels show the corresponding marginal distributions for $m_1$ and $m_2$, respectively, with the black curves indicating the total population and the grey bands representing the 90\% posterior predictive intervals. The underlying four-component sub-populations are shown as the coloured curves in the marginal distributions. Coloured points in the $(m_1,m_2)$ plane show the $\pistroke$ reconstruction. 
The $\pistroke$ solution appears in the marginal distribution as colored histograms.
The figure is cropped above $70\,\Msun$ for clarity, although the inferred distributions extend to higher masses.
}
\label{fig:m1_m2}
\end{figure*}

We solve for the $\pistroke$ samples applying the framework introduced in \citet{pi-stroke2} to \GWTCfive. The resulting reconstruction in the $(\mone,\mtwo)$ plane is shown in Fig.~\ref{fig:m1_m2}. 
The $\pistroke$ samples are conspicuously clustered into four visually distinct regions, which we mark with different colours to guide the eye. Motivated by these clusters, we construct a parametric population mixture model with four components in the $(\mone,\mtwo)$ plane, assigning each component its own mass and spin distribution. We denote these components by $\popL$ (blue), $\popMH$ (orange), $\popMD$ (green), and $\popU$ (violet).

\paragraph{$\popL$}
The $\popL$ component appears as a highly localized cluster near $\mone\simeq\mtwo\simeq10\,\Msun$. Following~\citet{Failedsn_Katerina}, we model this component with a truncated normal (\tn) distribution in $\mone$ and a \tn{} distribution in $q=\mtwo/\mone$.
The 25\% peak-density level for this model is indicated in Fig.~\ref{fig:m1_m2} with solid blue.
A prototypical event from this component is GW170608; $m_1 \approx 12 M_\odot, m_2 \approx 7 M_\odot$~\citep{GW170608}.

\paragraph{$\popMH$}
The $\popMH$ component appears as a horizontal line in the $(m_1, m_2)$ plane, extending from $m_1 \approx 20 M_\odot$ to $m_1 \approx 38 M_\odot$ at nearly-approximate $m_2\simeq10\,\Msun$. 
We model this population using a power-law distribution in $\mone$, together with a \tn{} distribution in $\mtwo$. This form is designed to capture binaries in which a low-mass secondary is paired with a heavier primary.
The 25\% peak-density level for this model is indicated in Fig.~\ref{fig:m1_m2} with dashed orange.
A prototypical event from this component is GW190412; $m_1 \approx 30 M_\odot, m_2 \approx 8 M_\odot$~\citep{GW190412}.

\paragraph{$\popMD$}
The $\popMD$ component appears as a line of slope $\approx$ 1 in the $(m_1, m_2)$ plane.
It extends from $m_1 \approx 18 M_\odot$ to $m_1 \approx 40 M_\odot$.
It may contain an excess near the $30$--$35\,\Msun$~\citep{Talbot_2018,mid30}. We model it with a mixture of a power-law distribution and a \tn{} in $\mone$ and a \tn{} distribution in $q$.
The 25\% peak-density level for this model is indicated in Fig.~\ref{fig:m1_m2} with dash-dot green.
A prototypical event from this component is the iconic GW150914; $m_1 \approx 36 M_\odot, m_2 \approx 29 M_\odot$~\citep{GW150914_properties}.

\paragraph{$\popU$}
The $\popU$ component captures the high-mass population; $m_1 \gtrsim 47 M_\odot$. We model this component with a power-law distribution in $\mone$ and a \tn{} distribution in $q$.
The 25\% peak-density level for this model is indicated in Fig.~\ref{fig:m1_m2} with dotted violet.
A prototypical event from this component is the iconic GW170729; $m_1 \approx 51 M_\odot, m_2 \approx 34 M_\odot$~\citep{GW170729_properties}.\footnote{It is interesting that all four prototype events were observed before 2020 in the GWTC-2 catalog \citep{gwtc-2}. It is only recently becoming apparent that they may embody four different formation channels.}

The total probability density (taking into account all four sub-populations) is shown in Fig.~\ref{fig:m1_m2} with black shading.
The full functional forms of the mass model and the corresponding parameter list are given in Appendix~\ref{app:model_details}.

For each sub-population $k$, we model the effective inspiral spin parameter with a mixture model:
\begin{align}
\pi_k(\chieff)
= &
(1-\xi_{\chi}^{(k)})\,
\tn(\chieff | \mu_\chi,\sigma_\chi) + \xi_{\chi}^{(k)}\,
\mathcal{U}(\chieff),
\label{eq:eff_spin_model}
\end{align}
Here, $\xi_{\chi}^{(k)}$ is the mixing fraction, which controls the relative contribution of a truncated normal and uniform distribution on the interval\footnote{The interval follows the expected $\chieff$ support for isotropic $1{\rm G}+2{\rm G}$ mergers with remnant spin $\tilde{a}\simeq0.69$ and $m_1\simeq2m_2$~\citep{antonini_U_chi_eff}.} $(-0.47, 0.47)$.
The truncated normal is hyper-parameterized with mean and variance $\mu_\chi, \sigma_\chi^{2}$. 
The truncated normal distribution is designed to model binaries with small spin magnitudes, including binaries formed in the field and also first-generation mergers in clusters.
The uniform distribution on the other hand is designed to model binaries with large black hole spins as expected for binaries with a second-generation black hole~\citep{antonini_U_chi_eff,hui_chi_eff_gp}. 

We model the spin-precession parameter $\chip$ with a shared \tn{} distribution with mean $\mu_P$ and variance $\sigma_P^2$.
The distributions of the other spin degrees of freedom are inherited by the LVK priors used for initial sampling.

It turns out that we gain additional insights using a spin model with physical coordinates. Thus, following~\cite{spin}, we also fit a physical-spin model in terms of $\chi_1,\chi_2,\cos\theta_1,\cos\theta_2$. We model the spin magnitudes in each subpopulation using truncated-normal distributions, with identical spin-magnitude distributions for the primary and secondary black holes within a given subpopulation.
The spin--orbit tilt angles are modeled as
\begin{align}
\pi_k(\cos\theta)=
(1-\xi_{{\rm tilt}}^{(k)})\,\frac{1}{2}
+\xi_{{\rm tilt}}^{(k)}
\tn\!\left(
\cos\theta \mid
\mu_{\rm tilt}=1,
\sigma_{\rm tilt}
\right),
\label{eq:tilt_model}
\end{align}
where the same tilt distribution is used for the primary and secondary black holes within each subpopulation. The mixing fraction $\xi_{\rm tilt}^{(k)}$ is allowed to vary between subpopulations, while the width $\sigma_{\rm tilt}$ is shared across all subpopulations. The parameter $\xi_{\rm tilt}^{(k)}$ controls the relative contribution of an isotropic component, uniform in $\cos\theta$ on $(-1,1)$, and an aligned component, modeled as a truncated normal with mean $\mu_{\rm tilt}=1$ and variance $\sigma_\text{tilt}^2$.

The redshift evolution is modeled with a power-law:
\begin{equation}
\pi(z)
\propto
\frac{\dd V_c}{\dd z}
(1+z)^{\kappa-1}.
\end{equation}
We do not impose a pair-instability gap in the secondary-mass distribution as a feature of the baseline model, in order to prevent it from driving the inferred Bayes factors, though our results are consistent with the existence of a gap~\citep{hui_gap}.

\section{Results}
\label{sec:results}

Table~\ref{tab:population_fractions} reports the inferred mixture fractions. The intrinsic \bbh{} population is dominated by the low-mass component, $\popL$, which accounts for approximately $70\%$ of the merger rate, in good agreement with previous analyses~\citep{Failedsn_Katerina,galaudage_2026}. The $\popMH$ and $\popMD$ components each contribute at the $\sim10$--$15\%$ level, while the upper-mass component $\popU$ contributes at the percent level. Thus, although most mergers belong to the low-mass population, the data support additional substructure associated with distinct regions of the mass distribution.
(This is thanks to selection effects, which make it comparatively easier to see more massive binaries.)

\begin{table}[t]
\centering
\caption{
Inferred mixture fractions for the four subpopulations. We report the median and 90\% credible interval for each component.
}
\label{tab:population_fractions}
\renewcommand{\arraystretch}{1.05}
\begin{tabular}{l c}
\hline
\hline
Component & Mixture fraction \\
\hline
$\popL$ & $0.709^{+0.069}_{-0.099}$ \\
$\popMH$ & $0.130^{+0.082}_{-0.059}$ \\
$\popMD$  & $0.141^{+0.079}_{-0.047}$ \\
$\popU$ & $0.014^{+0.013}_{-0.007}$ \\
\hline
\end{tabular}
\renewcommand{\arraystretch}{1.0}
\end{table}

Figure~\ref{fig:m1_m2} shows the inferred two-dimensional component-mass distribution in shaded black. 
The posterior predictive distribution closely follows the structures visible in $\pistroke$. The colored two-dimensional contours show 25\% peak-density levels for each subpopulation and serve to highlight the morphology of the distinct mass features.

The $\popL$ component is described by a \tn{} in $\mone$ with mean\footnote{All credibility intervals are 90\% unless otherwise stated.} $\mu_{m_1}^{\popL}=10.1^{+1.5}_{-0.8}\,\Msun$, together with a \tn{} in mass 
ratio with mean $\mu_q^{\popL}=0.85^{+0.12}_{-0.10}$. Above this component, we find a deficit of probability around $14.2^{+1.7}_{-3.1}\,\Msun$, with an estimated width of $2.7^{+6.9}_{-1.8}\,\Msun$ (see Appendix~\ref{app:gap}). This feature is consistent with the gap reported by~\citet{Failedsn_Katerina,gennari2026}, and attributed to the transition between successful and failed supernovae~\citep{Failedsn1}. The deficit is followed by a small feature around $18$–$20\,\Msun$, associated with the rising lower-mass edge of the $\popMH$ component, similar to the substructure reported in~\citet{Edelman_2023,Toubiana_2023,Heinzel_2025,tiwari2025pop,Farah_subpopulation,mould2026joint,galaudage_2026}. Importantly, the gap is not imposed by the prior: the lower boundary of the intermediate components is allowed to extend into the $10\,\Msun$ peak.
This increases our confidence that this low-mass gap may be real, although the data do not require it.

The two $\popMH$ and $\popMD$ components overlap substantially in primary mass, but differ strongly in their secondary-mass and mass-ratio structure. The $\popMH$ component forms a horizontal branch, with $\mtwo$ concentrated near the low-mass $10\,\Msun$ peak, whereas the $\popMD$ component forms a diagonal branch where $m_2 \approx m_1$. This highlights that the population structure cannot be fully captured by modelling the one-dimensional primary-mass distribution alone, consistent with the findings of \citet{Sadiq_2024}. 
We further note that the two components have different behaviour near their lower-mass edges.
On the interval $m_1 \in (16 M_\odot,20 M_\odot)$, the median integrated probability assigned to $\popMH$ exceeds that assigned to $\popMD$ by a factor of $\sim 4.9$. 
That is, the \popMH{} component may have a pile-up at $\approx 18\,M_\odot$.

The upper edge of \popMH{} is $m_1=37^{+20}_{-12}\,\Msun$; the upper edge of the \popMD{} is $m_1=55^{+13}_{-11}\,\Msun$; and the lower edge of \popU{} is $m_1=50^{+6}_{-12}\,\Msun$.
These edges are all consistent with the lower edge of the pair instability mass gap identified in \cite{antonini_U_chi_eff,Ray_2026,Mould_2026_pisn,hui_gap}: $44^{+5}_{-4} M_\odot$. 

\begin{figure}[t]
\centering
\includegraphics[width=0.95\columnwidth]{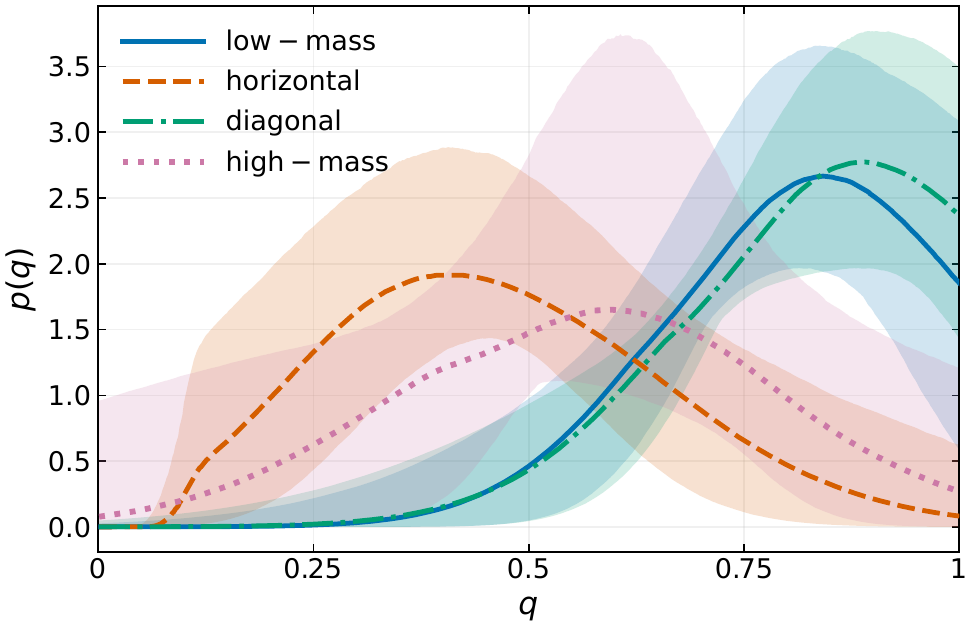}
\vspace{0.05cm}
\includegraphics[width=\columnwidth]{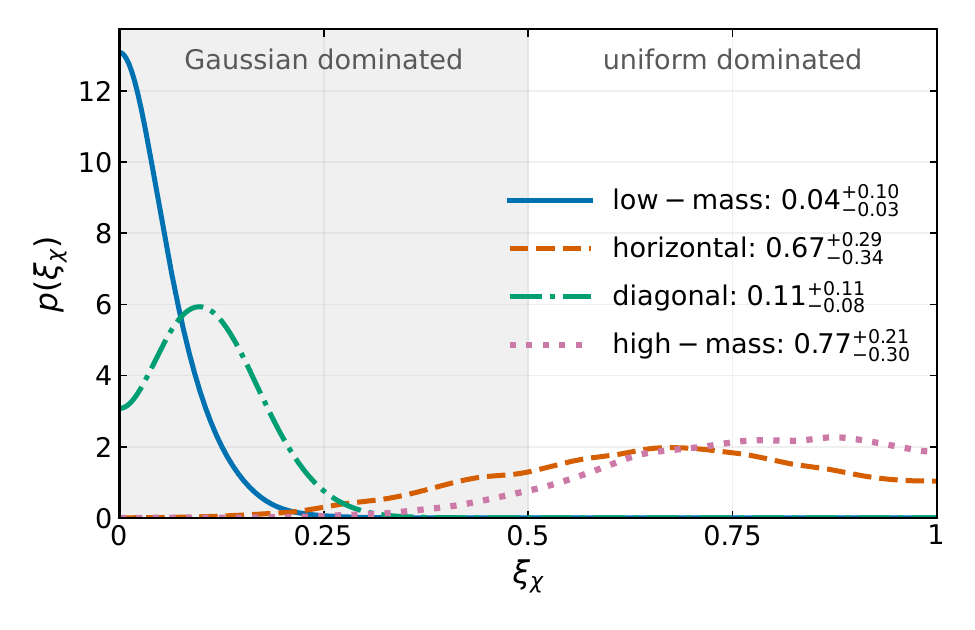}
\caption{
Inferred mass-ratio and effective-spin properties of the four reconstructed subpopulations.
Top: posterior predictive distributions for the mass ratio $q$, with shaded bands showing the corresponding 90\% posterior predictive intervals. The $\popL$ component and $\popMD$ component favour nearly equal masses, while $\popMH$ and $\popU$ prefer broader and more unequal-mass distributions. Bottom: posterior distributions for the spin-model mixture parameter $\xi_\chi$ (see Eq.~\eqref{eq:eff_spin_model}), which controls the relative contribution of the broad uniform component in the $\chieff$ distribution. Small values of $\xi_\chi$ correspond to a narrow distribution near $\chieff\simeq0$, while large values correspond to a broader effective-spin distribution.
}
\label{fig:xi_q}
\end{figure}

Figure~\ref{fig:xi_q} shows the inferred mass-ratio and $\chi_\text{eff}$ structures. The mass-ratio distributions reflect the morphology seen in Fig.~\ref{fig:m1_m2}: $\popL$ and $\popMD$ favour nearly equal masses, while $\popMH$ and $\popU$ prefer broader and more unequal-mass distributions. 

Turning our attention to the lower panel in Fig.~\ref{fig:xi_q}, we see that both the $\popL$ component and the $\popMD$ component prefer small $\xi_\chi$, characteristic of $\chieff$ distributions concentrated near zero. By contrast, $\popU$ favours the broad component, while $\popMH$ lies in between these regimes. This suggests that $\popMH$ has mixed spin properties, rather than resembling either the low-mass, nearly equal-mass components or the high-mass component alone. These findings are broadly consistent with those of~\citet{Sridhar:2025kvi,Ray_2026a,Banagiri_GWTC4,Farah_subpopulation,galaudage_2026}, who identify structure consistent with multiple distinct subpopulations.

\begin{figure*}[t]
\centering
\includegraphics[width=1.95\columnwidth]{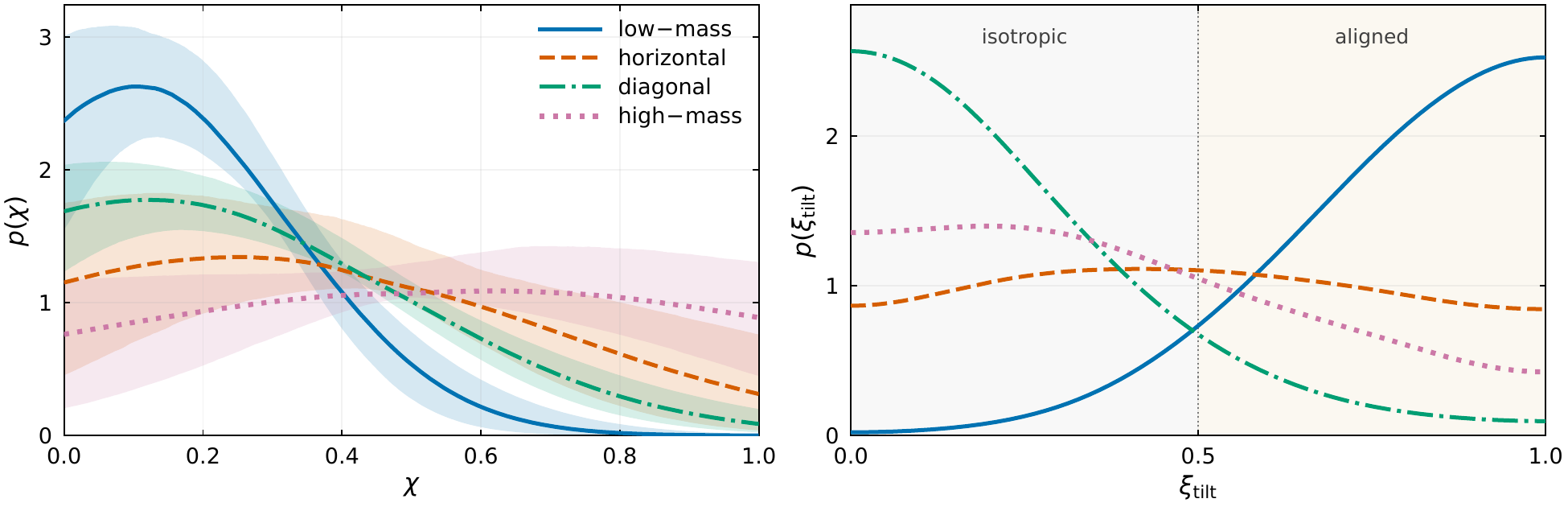}
\caption{
Inferred physical-spin properties of the four reconstructed subpopulations.
Left: posterior predictive distributions for the spin magnitude $\chi$, assumed to be identical for the primary and secondary black holes within each subpopulation. Shaded bands show the corresponding 90\% posterior predictive intervals.
Right: posterior distributions for the tilt-alignment mixing fractions \(\xi_{\rm tilt}^{(k)}\) defined in Eq.~\eqref{eq:tilt_model}. The parameter \(\xi_{\rm tilt}^{(k)}\) controls the relative contributions of the isotropic and preferentially aligned components.
The subpopulations exhibit distinct spin behaviour: $\popL$ favours low spin magnitudes and stronger alignment, $\popMD$ has low-to-moderate spins and weaker alignment, $\popMH$ extends to larger spin magnitudes with an intermediate alignment fraction, and $\popU$ has the broadest spin-magnitude distribution and broad support in the tilt-alignment fraction.}

\label{fig:spin_tilt}
\end{figure*}

Figure~\ref{fig:spin_tilt} shows the inferred physical-spin distributions. 
The $\popL$ component favours relatively small spin magnitudes and preferentially aligned tilts. The $\popMD$ component also prefers low-to-moderate spin magnitudes.
The $\popMD$ component \textit{may} exhibit broader tilt support, which would indicate that its low effective spins do not imply the same aligned-spin structure as $\popL$, but the error bars are not small enough to say confidently if the $\popL$ and $\popMD$ spin distributions are different.

The $\popMH$ component shows broader spin-magnitude support. Its spin-magnitude distribution has a mean $\mu_{\chi}^{\popMH}=0.24^{+0.29}_{-0.21}$ and width $\sigma_{\chi}^{\popMH}=0.45^{+0.22}_{-0.19}$. Its tilt distribution has an intermediate structure, consistent with the mixed behaviour inferred from $\chieff$. Finally, the upper-mass component $\popU$ has the broadest spin-magnitude support, with mean $\mu_{\chi}^{\popU}=0.58^{+0.36}_{-0.45}$ and width $\sigma_{\chi}^{\popU}=0.70^{+0.27}_{-0.36}$, and favours less aligned spin orientations. This is qualitatively consistent with a high-mass population containing hierarchical mergers.
This spin-magnitude hierarchy is broadly consistent with \citet{Galaudage_2025, galaudage_2026, berti_2026,wolfe2026}, who found smaller spins at lower masses and broader support for large spins above $\sim40\,M_\odot$.
The fact that the mass-defined subpopulations also exhibit distinct spin properties suggests that the different clusters may be tracing physically distinct formation channels.

\section{Model comparison}
\label{sec:model_comparison}

We compare the four-subpopulation model against standard parametric descriptions. For the physical-spin analysis, we compare against the LVK default population model~\citep{gwtc-5_population}, including the \texttt{Broken Power Law + 2 Peaks} mass model and the corresponding LVK redshift and spin prescriptions. The four-subpopulation model is preferred by a log Bayes factor of $\ln\mathcal{B}\simeq19.2$, indicating strong support for additional mass and spin substructure.

For the effective-spin analysis, we first benchmark several choices for the spin distribution in the LVK-like baseline model before comparing it to the four-subpopulation model. This step is intended to ensure that any preference for substructure is not driven by an overly restrictive spin prescription. 
We compare three prescriptions for the baseline spin distribution: independent \tn{} distributions in $\chieff$ and $\chip$; a skew-normal distribution for $\chieff$ combined with a \tn{} distribution for $\chip$~\citep{Banagiri_chi_eff}; and a \tn{} plus restricted-uniform model for $\chieff$, analogous to Eq.~\ref{eq:eff_spin_model}, combined with a \tn{} distribution for $\chip$. 
The skew-normal model is mildly preferred over the independent-\tn{} model, but the \tn{} plus restricted-uniform model is preferred over the skew-normal model by $\ln\mathcal{B}=4.9$. We therefore adopt the \tn{} plus restricted-uniform prescription as our baseline effective-spin model.
Using this baseline spin prescription, the four-subpopulation model is preferred over the corresponding LVK-like baseline by $\ln\mathcal{B}\simeq16.5$.

The consistency between the effective-spin and physical-spin analyses shows that the preference for substructure is not tied to a particular spin parameterization.

Finally, we explore an additional variant of the physical-spin model in which the redshift evolution varies independently between subpopulations. This variant leads to only modest evidence differences, with $|\ln\mathcal{B}|\lesssim0.5$, indicating that the data do not require separate redshift-evolution parameters for the different subpopulations. This contrasts with the findings of~\citet{Sylvia_subpopulation}, who find that the peak near~$10\,\Msun$ evolves more rapidly with redshift. We therefore adopt the models described in Sec.~\ref{sec:data_model} as our baseline analyses.

\section{Astrophysical interpretation}
The $\popL$ component is centred at $m_1\simeq10\,\Msun$. Its mass scale, mass-ratio distribution, and low-spin properties are broadly consistent with expectations for an isolated binary-formation channel~\citep{isolated_channel_review,CE,Fuller_2019}. A striking feature of this component is that it is separated from the heavier subpopulations by a gap at approximately $12$–$16\,\Msun$, even though no such gap is imposed explicitly in our model. Similar structure was first identified in the chirp-mass distribution~\citep{Tiwari_2021,Tiwari_2025,chirp_mass_gap}, and has been interpreted in terms of compactness-driven stellar evolution~\citep{Schneider_2021,Schneider_2023,Galaudage_2025,galaudage_2026} and failed-supernova/direct-collapse scenarios~\citep{Failedsn1,Failedsn_Katerina}. Our results are broadly consistent with these interpretations.
In these pictures, a narrow range of progenitor core structures preferentially produces black holes near $10\,\Msun$, while adjacent progenitor masses are less likely to populate this black-hole mass range, producing a deficit before the higher-mass population turns on. Quantitatively, we find that the rate decreases by a factor of approximately 26 across $12$--$16\,\Msun$ relative to the LVK default model (see Appendix~\ref{app:gap}). We therefore interpret the $\popL$ component as consistent with this class of isolated stellar-evolution scenarios and as distinct from the heavier components above the gap, although our analysis does not distinguish between the different physical mechanisms.

The $\popMD$ component peaks at $31.0^{+2.7}_{-4.2}\,\Msun$ and favours nearly equal masses. Its small $\chieff$ values suggest that this component is dominated by first-generation (1G) black holes rather than hierarchical merger remnants. The main question is therefore whether these 1G binaries are formed predominantly through isolated binary evolution or in dense stellar environments.

The physical-spin analysis allows for broad tilt support, close to the expectation for randomly oriented spins and different from the more aligned $\popL$ component. However, it is difficult to draw firm conclusions about the spin orientation because the statistical significance for broad tilts in $\popMD$ is only marginal. Nevertheless, the results are consistent with the effective-spin distribution, because small spin magnitudes combined with broad spin orientations can still produce a narrow $\chieff$ distribution centred near zero~\citep{callister_chi_eff}.

The mass distribution provides an additional clue. Rather than turning on sharply above the failed-supernova gap as a smooth high-mass extension of $\popL$, the $\popMD$ population has relatively little support near $m_1\sim20\,\Msun$ and rises toward the $30$--$35\,\Msun$ region. Quantitatively, we estimate that the abundance of $\popMD$ near $m_1\simeq32\,\Msun$ exceeds that near $m_1\simeq20\,\Msun$ by a median factor of about $2.5$. This behaviour is driven by the localized \tn{} component, which contributes $78^{+20}_{-34}\%$ of $\popMD$, compared with the smoother power-law component. The low support of $\popMD$ near $m_1\sim20\,\Msun$ is not caused by a general absence of black holes in this mass range: the $\popMH$ component has substantial support there, exceeding $\popMD$ by a median factor of about $4.9$.

This rising intermediate-mass structure resembles the broad mass features produced in globular-cluster simulations, which can generate an excess of first-generation \bbh{} mergers around this mass scale~\citep{oconnor_clutsers}.
Together with the possible broad tilt support, we mildly favour an interpretation in which $\popMD$ includes a contribution from 1G \bbh{} formation in dense stellar environments, such as globular clusters~\citep{Antonini_clusters,Ray_2025}. However, we do not regard this as a smoking-gun identification of the formation channel. An isolated-binary origin remains entirely plausible, since isolated evolution can also produce massive, low-effective-spin \bbh{} systems in this mass range~\citep{Belczynski_2020}.

The $\popMD$ component also naturally connects to the well-studied excess near $35\,\Msun$~\citep[e.g.,][]{Talbot_2018,mid30}. In our reconstruction, this feature appears not as an isolated narrow peak, but as the main part of a broader intermediate-mass, near-equal-mass subpopulation.

%C3
The $\popMH$ component is the most difficult to interpret. It corresponds to an unequal-mass population in which the secondary mass is close to the $10\,\Msun$ scale of the $\popL$ component, while the primary extends above the failed-supernova gap. Its physical-spin distribution is broader than that of $\popL$ and $\popMD$, extending to larger spin magnitudes, but remains below the characteristic spin scale of $\popU$. The typical spin magnitude is $\chi\sim0.25$. This is consistent with the effective-spin analysis, which favours a mixture of support near $\chieff\simeq0$ and a broader contribution. Moreover,~\citet{Li_2025,pix_pop_gwtc5,rinaldi2026,flanagan2026_2,flanagan2026,Li_2026} report a mild preference for positive $\chieff$ in a similar mass range, suggesting that at least part of this population may retain some spin–orbit alignment. Thus, $\popMH$ does not have the clean signature of either a standard low-spin isolated-binary population or a purely dynamical population. This is consistent with the results of~\citet{flanagan2026,galaudage_2026}. 

A key clue is the secondary-mass distribution. The inferred secondary-mass scale of $\popMH$ is consistent with the $10\,\Msun$ peak in $\popL$, with $9.2^{+2.3}_{-2.4}\,\Msun$ compared to $10.1^{+1.5}_{-0.8}$, although it is somewhat broader, with $\sigma=3.8^{+1.8}_{-1.7}$ compared to $\sigma=1.7^{+1.7}_{-0.7}$. This broadening may partly reflect measurement uncertainty, since the secondary mass is typically less well measured in unequal-mass systems. We therefore interpret $\popMH$ as a population in which a black hole associated with the low-mass failed-supernova/direct-collapse peak is paired with a heavier primary black hole.

Such unequal-mass pairings are astrophysically non-trivial. In isolated binary evolution, their formation depends on the stability of mass transfer, common-envelope survival, natal kicks, and whether the binary remains bound through compact-object formation~\citep[e.g.][]{Hills,Tauris1998,Holgado_2019,Marchant_2021}. The inferred primary-mass spectrum of $\popMH$ is steep, with $\alpha_{\popMH}=5.2^{+4.0}_{-5.7}$ under the convention $p(m)\propto m^{-\alpha}$, much steeper than the Salpeter value $\alpha\simeq2.35$~\citep{Salpeter}. This may indicate that the efficiency of producing and retaining such unequal-mass systems decreases rapidly toward larger primary masses. The upper mass scale of $\popMH$ may therefore reflect not only the underlying black-hole mass spectrum, but also the declining survival efficiency of increasingly unequal-mass binaries.

One plausible interpretation is stable-mass transfer in isolated binary evolution. In this scenario, mass transfer can reverse the mass ratio, while later tidal interactions spin up the progenitor of the second-born black hole before collapse. This provides a possible route to unequal-mass binary black holes with non-negligible or positive $\chieff$, and could explain the broader spin support inferred for $\popMH$. Natal kicks during compact-object formation can further tilt the orbital plane relative to the component spins, broadening the $\chieff$ distribution and potentially producing systems with negative $\chieff$ if the tilt is sufficiently large~\citep{uneqal_mass_spunup}. However, this interpretation depends on uncertain binary-evolution physics. Population-synthesis studies find that systems combining large mass asymmetry with substantial spin can be rare under standard assumptions; for example,~\citet{Zevin_2022} argue that GW190412-like systems are difficult to produce through mass-ratio reversal and tidal spin-up unless highly super-Eddington accretion or less efficient angular-momentum transport is invoked.

An alternative possibility is suggested by~\citet{willcox2025}, who use the COMPAS binary-population-synthesis
code~\citep{Compas2022,compas_2} to identify an isolated-binary formation
pathway in which strong winds during the luminous-blue-variable phase
strip the envelope of the initially more massive star, delaying binary
interaction until after the first-born BH has formed. The subsequent
interaction typically leads to common-envelope evolution. This pathway
produces binaries containing a $\sim10\,\Msun$ BH paired with a
heavier companion. 
A qualitatively similar phenomenological cross-gap population was obtained by~\citet{Failedsn1} through the pairing of BHs from below and above the failed-supernova gap.
However, whether either model can also reproduce the
spin behaviour and steep primary-mass decline inferred for $\popMH$
remains unclear. We therefore regard isolated-binary evolution as a
plausible interpretation of $\popMH$, either through stable mass transfer or the luminous-blue-variable-assisted common-envelope pathway, but not a unique one.

% C4
The $\popU$ component has the clearest interpretation as a hierarchical-merger population. Its spin magnitude distribution has mean value around the expected remnant spin of a black-hole merger, $\chi\sim0.7$~\citep{Gerosa_2021}, strongly suggesting that this component contains second-generation (2G) black holes. The most natural configuration is therefore a 1G+2G binary, although a smaller contribution from 2G+2G systems is also possible. A dominant 2G+2G contribution would be less likely, since it requires both merger remnants to be retained and paired again.

This interpretation is also supported by the mass-ratio distribution. The inferred distribution is broad and peaks at $q=0.58^{+0.19}_{-0.25}$, as expected if a massive merger remnant is paired with a lower-mass 1G black hole~\citep{Gerosa_2021}. The spin information points to the same conclusion. Both the $\chieff$ distribution and the physical-spin analysis are consistent with broad spin–orbit misalignment, as expected for dynamically assembled binaries in dense stellar environments, where repeated mergers can occur and the spin directions need not be correlated with the binary orbital angular momentum~\citep{Gerosa_2021,Antonini_clusters}.

Table~\ref{tab:subpopulation_summary} gives a compact summary of these results and their interpretation.
\begin{table*}[t]
\centering
\caption{Qualitative summary of the inferred subpopulations.}
\label{tab:subpopulation_summary}
\begin{tabular}{l c c c c}
\hline
Comp. & $m_1$ & $m_2$ & Spins & Interpretation \\
\hline
$\popL$ &
$\sim10\,\Msun$ &
$\sim8\,\Msun$ &
low, aligned &
low-mass failed SN / isolated \\

$\popMH$ &
$20$–$40\,\Msun$ &
$\sim10\,\Msun$ &
mixed &
stable MT / cross-gap pairing \\

$\popMD$ &
$20$–$50\,\Msun$ &
$q\sim1$ &
low, isotropic &
dense environments / isolated \\

$\popU$ &
high &
broad / low $q$ &
high, isotropic &
hierarchical mergers \\
\hline
\end{tabular}
\end{table*}

\section{Conclusions}

We present a four-component description of the binary black-hole population, combining data-driven reconstruction with parametric population inference. Our results suggest that the population contains multiple structures associated with distinct mass scales and spin properties.

The $\popL$ component is centred near $10\,\Msun$ and is associated with the low-mass failed-supernova/direct-collapse peak. The $\popMD$ component peaks near $30$–$35\,\Msun$ and favours nearly equal masses. Its low effective spins arise from small spin magnitudes together with broad spin–orbit tilt support, rather than from strong alignment, suggesting an origin in dense stellar environments or another channel distinct from a smooth continuation of the low-mass population. The $\popMH$ component pairs a secondary near the $10\,\Msun$ peak with a heavier primary, and may reflect an unequal-mass isolated-binary channel involving either stable mass transfer and spin-up or luminous-blue-variable-assisted common-envelope evolution. Finally, the $\popU$ component has the clearest hierarchical-merger signature, with large spin magnitudes, broad mass ratios, and less aligned spin orientations.

Overall, the reconstructed subpopulations align with physically motivated mass scales associated with black-hole formation. The low-mass component, $\popL$, occupies the $\sim10\,\Msun$ failed-supernova/direct-collapse peak, while $\popMH$ and $\popMD$ emerge above the depleted region that separates this peak from the higher-mass population and remain mostly below the pair-instability regime. The highest-mass component, $\popU$, begins near the onset of this regime and extends to higher masses. Since both the failed-supernova and pair-instability features are physical mass scales in black-hole formation, their appearance across multiple reconstructed subpopulations suggests that the decomposition is not merely fitting arbitrary structure, but is recovering physically meaningful features associated with compact-object formation.

A recent study by Ray \textit{et al.}\ (in preparation) independently identified similar subpopulation structure in the BBH population using a different non-parametric approach. Although its decomposition into subpopulations differs from that presented here, the overall results are in agreement with those reported in this study, providing independent support for the inferred population structure.

More generally, we note that the subpopulation interpretation is not unique. For example, \citet{tiwari2026chirpmass} proposed that some of the structure in the \GWTCfive{} chirp-mass distribution may instead reflect a hierarchical-merger mass ladder, with different features corresponding to different merger generations.

Larger gravitational-wave catalogues will test whether these structures persist and will help distinguish between the formation-channel interpretations proposed here.

\section*{Acknowledgments}

This work is supported through the Australian Research Council (ARC) Centre of Excellence CE230100016, Discovery Projects DP220101610 and DP230103088, and LIEF Project LE210100002.
This material is based upon work supported by NSF's LIGO Laboratory which is a major facility fully funded by the National Science Foundation.
The authors are grateful for computational resources provided by the LIGO Laboratory and supported by National Science Foundation Grants PHY-0757058 and PHY-0823459.
LIGO Laboratory and Advanced LIGO are funded by the United States National Science Foundation (NSF) as well as the Science and Technology Facilities Council (STFC) of the United Kingdom, the Max-Planck-Society (MPS), and the State of Niedersachsen/Germany in support of the construction of Advanced LIGO and construction and operation of the GEO600 detector. Additional support for Advanced LIGO was provided by the Australian Research Council. Virgo is funded, through the European Gravitational Observatory (EGO), by the French Centre National de Recherche Scientifique (CNRS), the Italian Istituto Nazionale di Fisica Nucleare (INFN) and the Dutch Nikhef, with contributions by institutions from Belgium, Germany, Greece, Hungary, Ireland, Japan, Monaco, Poland, Portugal, Spain. The construction and operation of KAGRA are funded by Ministry of Education, Culture, Sports, Science and Technology (MEXT), and Japan Society for the Promotion of Science (JSPS), National Research Foundation (NRF) and Ministry of Science and ICT (MSIT) in Korea, Academia Sinica (AS) and the Ministry of Science and Technology (MoST) in Taiwan.

\appendix

\section{Population-model details}
\label{app:model_details}

Here we provide the full functional form of the mass model used in Sec.~\ref{sec:data_model}.
The population is modeled as a mixture of four subpopulations,
\begin{equation}
p(\mone,q,z,\bm{\chi}\mid\Lambda) = \\
p_k(z\mid\Lambda)\sum_{k=1}^{4}
f_k\,
p_k(\mone,q\mid\Lambda_k)\,
p_k(\bm{\chi}\mid\Lambda_k)\,
,
\label{eq:population_mixture}
\end{equation}
where $f_k$ are mixture fractions satisfying $\sum_k f_k=1$, and $\bm{\chi}$ denotes the spin variables. 
%The population is modeled as the mixture in Eq.~\eqref{eq:population_mixture}. 
Unless otherwise stated, all normal distributions are truncated to the physical support of the corresponding variable. We denote a truncated normal distribution by \tn{} and a smoothed power law by $\mathcal{PL}$.
The effective-spin and physical-spin models contain 40 and 45 sampled hyperparameters, respectively; see Table~\ref{tab:appendix_parameter_list}. 
The prior ranges and posterior corner plots are available in the supporting GitHub repository.\footnote{\paperrepo}

\subsection{Mass model}
The low-mass component, $\popL$, is
\begin{equation}
\pL(\mone,q)
\propto
\tn
\!\left(\mone;\mu_{m_1}^{\popL},\sigma_{m_1}^{\popL}\right)
\,
\tn
\!\left(q;\mu_q^{\popL},\sigma_q^{\popL}\right).
\end{equation}
The horizontal intermediate-mass component, $\popMH$, is parameterized in $(\mone,\mtwo)$ as

\begin{equation}
p_{\popMH}(\mone,\mtwo) \propto 
\mathcal{PL}
\!\left(\mone;\alpha_{\popMH}\right)
\,
\tn
\!\left(\mtwo;\mu_{m_2}^{\popMH},\sigma_{m_2}^{\popMH}\right).
\end{equation}
The diagonal intermediate-mass component, $\popMD$, is

\begin{equation}
\pMD(\mone,q) \propto
(1-\lambda_{\popMD})\,
\mathcal{PL}\!\left(\mone;\alpha_{\popMD}\right) + 
\lambda_{\popMD}\,
\tn
\!\left(\mone;\mu_{\rm pk}^{\popMD},\sigma_{\rm pk}^{\popMD}\right)
\tn
\!\left(q;\mu_q^{\popMD},\sigma_q^{\popMD}\right),
\end{equation}
where $\lambda_\popMD$ is a mixing fraction between the power-law component and the truncated Gaussian.
The high-mass component, $\popU$, is
\begin{equation}
\pU(\mone,q)
\propto
\mathcal{PL}
\!\left(\mone;\alpha_{\popU}\right)
\,
\tn
\!\left(q;\mu_q^{\popU},\sigma_q^{\popU}\right).
\end{equation}

When the model component is evaluated in $(\mone,q)$, we include the Jacobian factor $\dd\mtwo/\dd q=\mone$.

\subsection{Smooth mass tapers}
\label{app:smoothing}

All mass components are restricted to finite support. To avoid sharp discontinuities at the component boundaries, we multiply the relevant mass distributions by Planck tapers. For a lower boundary $m_{\min}$ and smoothing scale $\Delta m$, we define
\begin{equation}
T_-(m;m_{\min},\Delta m)
=
\begin{cases}
0, & m\leq m_{\min},\\[3pt]
g(x), & m_{\min}<m<m_{\min}+\Delta m,\\[3pt]
1, & m\geq m_{\min}+\Delta m,
\end{cases}
\label{eq:planck_taper_low}
\end{equation}
where
$
x=\frac{m-m_{\min}}{\Delta m}
$, and  
$g(x)=
\left[
\exp\left(
\frac{1}{x}+\frac{1}{x-1}
\right)+1
\right]^{-1}$.
The corresponding upper-edge taper is
\begin{equation}
T_+(m;m_{\max},\Delta m) = T_-(m_{\max}-m;0,\Delta m),
\end{equation}
which is unity for $m\leq m_{\max}-\Delta m$ and smoothly falls to zero at $m=m_{\max}$. The full smoothing window is therefore

\begin{equation}
W(m;m_{\min},m_{\max},\Delta m_-,\Delta m_+) =
T_-(m;m_{\min},\Delta m_-)
T_+(m;m_{\max},\Delta m_+).
\end{equation}
Smoothed power laws are obtained by multiplying the power law by this window and normalizing over the allowed mass range.

\begin{table}[t]
\caption{
Grouped list of sampled hyperparameters for the population models. The effective-spin
and physical-spin models share the same mixture and mass model, but differ in their spin
parameterization. The fourth mixture fraction is fixed by $f_4=1-f_1-f_2-f_3$ and is not counted
as a sampled parameter.
}
\label{tab:appendix_parameter_list}
\begin{ruledtabular}
\begin{tabular}{l c c c}
\footnotesize
Block & Sampled hyperparameters & Effective spin & Physical spin \\
\hline

Mixture fractions
& $f_1,f_2,f_3$
& 3 & 3 \\

$\popL$ component 
& $m_{1,\min}^{\rm L}$, $m_{1,\max}^{\rm L}$, $\Delta m^{\rm L}$,
$\mu_{m_1}^{\rm L}$, $\sigma_{m_1}^{\rm L}$,
$\mu_q^{\rm L}$, $\sigma_q^{\rm L}$
& 7 & 7 \\

$\popMH$ component 
& $\alpha_{m_1}^{\rm MH}$, $\Delta m^{\rm MH}$,
$m_{1,\min}^{\rm MH}$, $m_{1,\max}^{\rm MH}$,
$\mu_{m_2}^{\rm MH}$, $\sigma_{m_2}^{\rm MH}$
& 6 & 6 \\

$\popMD$ component 
& $\alpha_{m_1}^{\rm MD}$, $\Delta m^{\rm MD}$,
$m_{1,\min}^{\rm MD}$, $m_{1,\max}^{\rm MD}$,
$\lambda_{\rm pk}^{\rm MD}$, $\mu_{\rm pk}^{\rm MD}$,
$\sigma_{\rm pk}^{\rm MD}$, $\mu_q^{\rm MD}$, $\sigma_q^{\rm MD}$
& 9 & 9 \\

$\popU$ component
& $\alpha_{m_1}^{\rm U}$, $\Delta m^{\rm U}$,
$m_{1,\min}^{\rm U}$, $m_{1,\max}^{\rm U}$,
$\mu_q^{\rm U}$, $\sigma_q^{\rm U}$
& 6 & 6 \\

Redshift evolution
& $\kappa$ 
& 1 & 1 \\

Effective-spin model
& $\mu_\chi$, $\sigma_\chi$, $\xi_1,\xi_2,\xi_3,\xi_4$,
$\mu_{\chip}$, $\sigma_{\chip}$
& 8 & -- \\

Physical-spin model
& For each component $k$:
$\mu_{\chi}^{(k)}$, $\sigma_{\chi}^{(k)}$,
$\xi_{\rm tilt}^{(k)}$;
with a common $\sigma_{\rm tilt}$
& -- & 13 \\

\hline
Total
& 
& 40 & 45 \\
\end{tabular}
\end{ruledtabular}
\end{table}

\subsection{Effective-spin variables}
\label{app:effective_spins}

For the effective-spin analysis, we describe each binary in terms of the two standard spin combinations $\chieff$ and $\chip$. These variables are functions of the component spin magnitudes, $\chi_1$ and $\chi_2$, and the spin--orbit tilt angles, $\theta_1$ and $\theta_2$. The effective inspiral spin is defined as~\citep{chi_eff}
\begin{equation}
\chieff =
\frac{\mone \chi_1\cos\theta_1+\mtwo \chi_2\cos\theta_2}
{\mone+\mtwo}.
\end{equation}
This parameter measures the mass-weighted projection of the component spins along the orbital angular momentum and is typically the best-measured spin combination in compact-binary observations.

We also use the effective precession spin parameter~\citep{chi_p},
\begin{equation}
\chip =
\max\left\{
\chi_1\sin\theta_1,\,
\frac{3+4q}{4+3q}\,q\,\chi_2\sin\theta_2
\right\},
\end{equation}
where $q=\mtwo/\mone\leq1$. This parameter captures the leading-order in-plane spin contribution that drives precession of the orbital plane.

\section{Gap Analyses}
\label{app:gap}

Here we quantify the low-mass feature by comparing our inferred primary-mass distribution to the LVK default population model using a moving-window probability ratio. 
For a window of width $\Delta m$ centred at $m_1$, we define
\begin{equation}
W_{m_1}
=
\left[m_1-\frac{\Delta m}{2},\,m_1+\frac{\Delta m}{2}\right],
\end{equation}
and compute the probability assigned to this window as
\begin{equation}
P_X(W_{m_1})
=
\int_{m_1-\Delta m/2}^{m_1+\Delta m/2}
p_X(m_1')\,d m_1',
\end{equation}
where $X$ denotes either our four-subpopulation model or the LVK default model, and $p_X(m_1)$ is the corresponding primary-mass distribution. We then define the moving-window probability ratio
\begin{equation}
R(m_1)
=
\frac{P_{\rm ours}(W_{m_1})}
     {P_{\rm LVK}(W_{m_1})}.
\end{equation}
Values $R<1$ indicate that our model assigns less probability than the LVK default model to that mass window. We use $\Delta m=0.5\,\Msun$ and scan the interval $10$--$23\,\Msun$.

\begin{figure}[t]
\centering
\includegraphics[width=0.49\columnwidth]{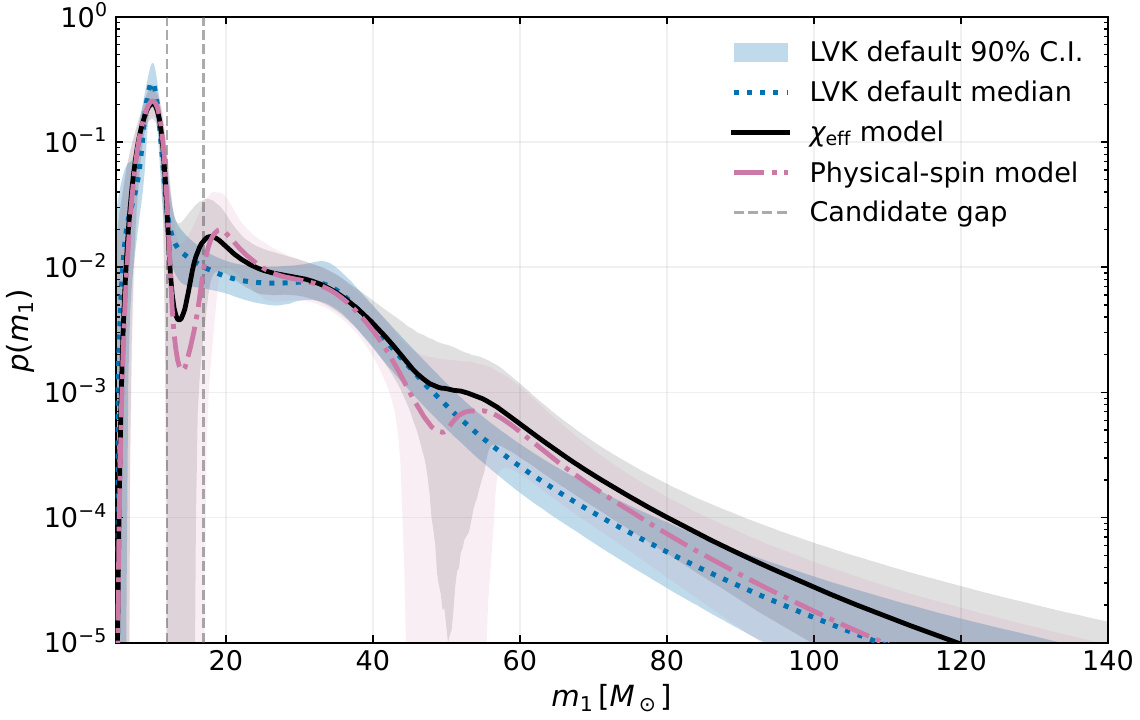}
%\vspace{0.05cm}
\includegraphics[width=0.49\columnwidth]{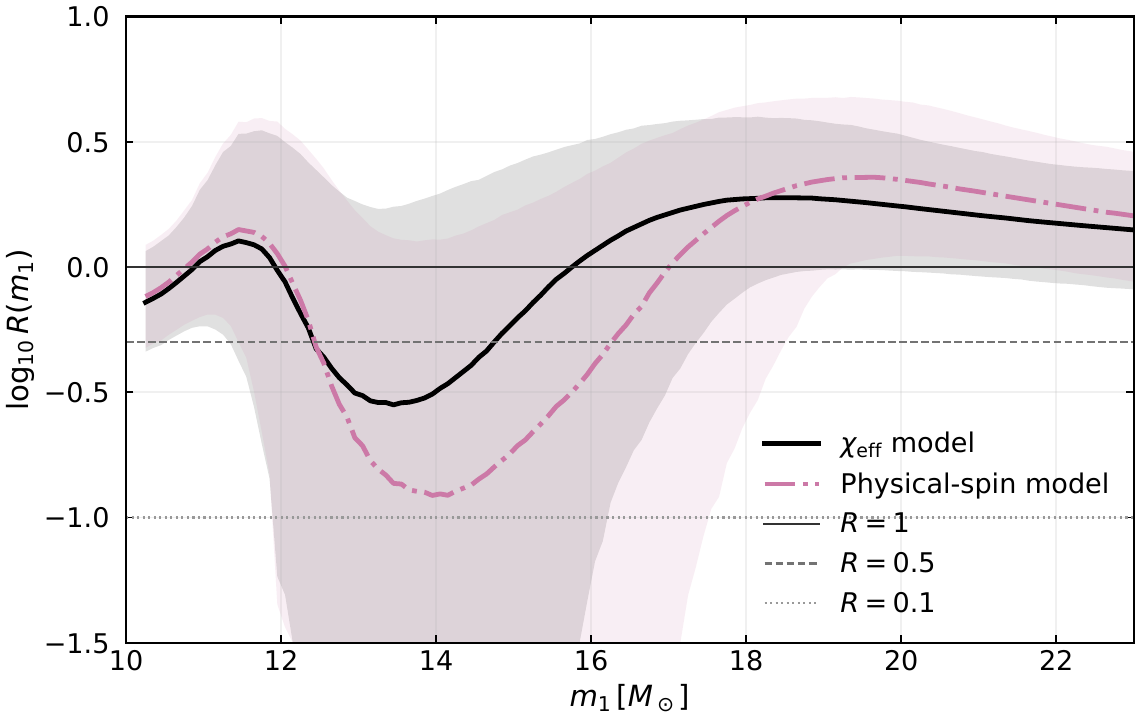}
\caption{
Low-mass depletion relative to the LVK default population model.
Left: posterior predictive distributions for the primary mass $m_1$ from the $\chi_{\rm eff}$ and physical-spin four-subpopulation models, compared with the LVK default model.
Right: moving-window probability ratios,
$R(m_1)=P_{\rm ours}(W_{m_1})/P_{\rm LVK}(W_{m_1})$, shown in terms of $\log_{10}R$ for the $\chi_{\rm eff}$ and physical-spin models.
Values below zero indicate mass windows where each four-subpopulation model assigns less probability than the LVK default model.
Both analyses show a localized minimum near $m_1\simeq14\,M_\odot$, indicating a depletion between the low-mass $10\,M_\odot$ peak and the higher-mass subpopulations.
}
\label{fig:gap}
\end{figure}

Figure~\ref{fig:gap} shows the comparison between our inferred primary-mass distribution and the LVK default model, together with the moving-window probability ratio. The $\log_{10}R(m_1)$ curve shows a localized depletion between the low-mass $10\,\Msun$ peak and the higher-mass subpopulations.

To quantify the location and width of the depletion, we perform a least-squares fit of a Gaussian dip to each posterior draw of the moving-window ratio,
\begin{equation}
\log_{10}R(m_1)=A-D\exp\left[
-\frac{(m_1-\mu)^2}{2\sigma^2}
\right].
\end{equation}
Here, $A$ is the local baseline, $D$ is the depth of the Gaussian depletion relative to that baseline, $\mu$ denotes the centre of the depletion and $\sigma$ its Gaussian width. We report the corresponding full width at half maximum (FWHM).
For the physical-spin model, the resulting fits give
\begin{equation}
\mu
=
14.2^{+1.7}_{-3.1}\,\Msun,
\qquad
{\rm FWHM}
=
2.7^{+6.9}_{-1.8}\,\Msun,
\end{equation}
where the uncertainties denote the 5th--95th percentile range. The fitted depletion depth has a long tail, because some posterior draws assign nearly zero probability to the depleted window. We therefore use the fitted location and width as a characterization of the depletion scale, rather than interpreting the depth as a precisely measured quantity. The fitted Gaussian dips give a characteristic minimum of
\begin{equation}
R_{\rm min}= 0.04^{+0.60}_{-0.04},
\end{equation}
corresponding to a depletion by a factor of $\sim26$ relative to the LVK default model.

\section{Mass distributions of the individual subpopulations}

\begin{figure}[t]
\centering
\includegraphics[width=1\columnwidth]{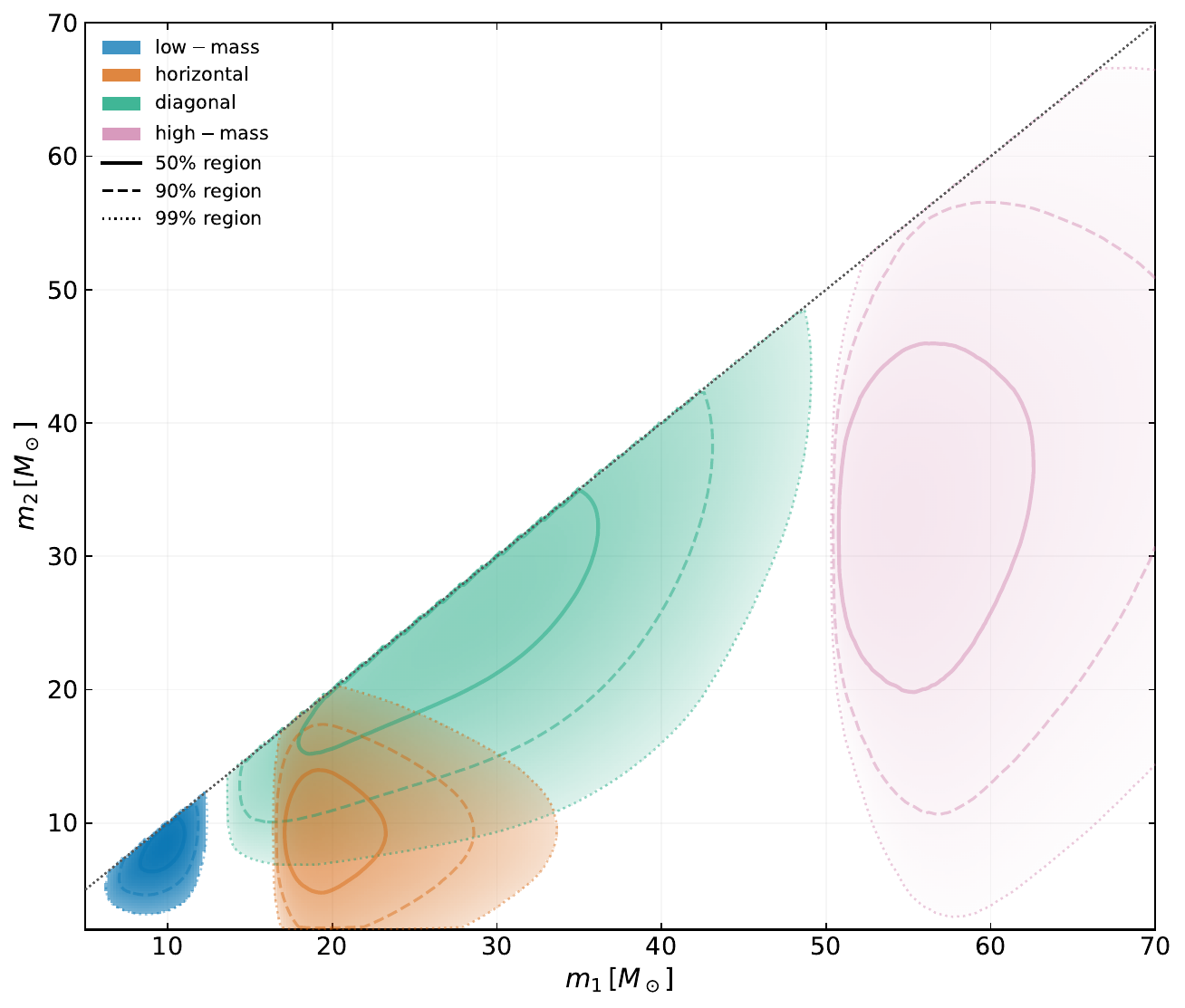}
\caption{
Mass distributions of the four inferred binary-black-hole subpopulations in the $(m_1,m_2)$ plane.
The coloured shading shows the median posterior-predictive density of each subpopulation, weighted by its inferred population fraction, so that the relative intensities indicate the components' contributions to the total population. For clarity, the shading is restricted to the highest-density region containing 99\% of the probability of each subpopulation within the plotted mass range.
For each subpopulation, the solid, dashed, and dotted contours enclose 50\%, 90\%, and 99\%, respectively, of the probability represented within the plotted mass range.
}

\label{fig:extra_m1m2}
\end{figure}

To further illustrate the inferred population, we present in Fig.~\ref{fig:extra_m1m2} the contributions from the individual subpopulations in the $(m_1,m_2)$ plane. This representation makes it easier to identify the regions of mass space occupied by each component and to visualise how they combine to form the full population.

\bibliographystyle{apsrev4-2}
\bibliography{ref}
\end{document}